# Evaluating the weight sensitivity in AHP-based flood risk estimation models

**Hongping Zhang**[a], **Zhenfeng Shao**[a, *], **Bin Hu**[a], **Xiao Huang**[b], **Jinqi Zhao**[a], **Wenfu Wu**[a], **Yewen Fan**[a]

[a] State Key Laboratory of Information Engineering in Surveying, Mapping and Remote Sensing, Wuhan University, Wuhan 430079, China;
yx_zhping@126.com (H.Z.); shaozhenfeng@whu.edu.cn (Z.S.); hubin763259288@whu.edu.cn (B.H.); masurq@whu.edu.cn (J. Z); wuwwf09140818@whu.edu.cn (W.W.) ; fyw@whu.edu.cn(W.F.).

[b] Department of Geosciences, University of Arkansas, Fayetteville, AR 72701, USA; xh010@uark.edu

[*] Corresponding author: shaozhenfeng@whu.edu.cn (Z.S.)

**Abstract:** In the analytic hierarchy process (AHP) based flood risk estimation models, it is widely acknowledged that different weighting criteria can lead to different results. In this study, we evaluated and discussed the sensitivity of flood risk estimation brought by judgment matrix definition by investigating the performance of pixel-based and sub-watershed-based AHP models. Taking a flood event that occurred in July 2020 in Chaohu basin, Anhui province, China, as a study case, we used the flood areas extracted from remote sensing images to construct ground truth for validation purposes. The results suggest that the performance of the pixel-based AHP model fluctuates intensively given different definitions of judgment matrixes, while the performance of sub-watershed-based AHP models fluctuates considerably less than that of the pixel-based AHP model. Specifically, sub-watershed delimitated via multiple flow direction (MFD) always achieves increases in the correct ratio and the fit ratio by >35% and >5% with the pixel-based AHP model, respectively.

*1. Introduction*

Floods are worldwide natural events, commonly occur in the interwoven areas of river networks, usually driven by extreme or continuous rainfall (Betancourt-Suárez et al., 2021; Majeed et al., 2021; Noaa, 2019). During a large flood event, rainwater converges from mountain areas, slope lands and then flows through rushing paths (e.g., valleys, stams, and rivers) at the watershed scale (Zhang et al., 2019). The excessive rainwater can overflow from expected storage spaces such as green roofs, retention ponds, reservoirs, lakes, streams, rivers, leading to flooded situations in these low-lying lands nearby rainwater rushing paths (Jahangir et al., 2019). Improving the accuracy of flood risk estimation in river network interwoven areas is able to support future flood risk management, thus benefiting the protection of the local economy.

The analytic hierarchy process (AHP) is a popular multi-criteria decision method that relies on expert knowledge(Liu et al., 2021). In an AHP-based flood risk estimation model, direct flood caused factors and surface runoff production characters are widely acknowledged as major influencing factors by many scholars (Bathrellos et al., 2017; Chen et al., 2021; Ekmekcioglu et al., 2021; Ghosh and Kar, 2018; Huu et al., 2020; Kanani-Sadat et al., 2019; Mishra and Sinha, 2020; Nachappa et al., 2020; Pham et al., 2021; Ribeiro Araujo Junior and Tavares Junior, 2020; Shahiri Tabarestani and Afzalimehr, 2021; Skilodimou et al., 2019). With the development of spatial information techniques, indicators involved in flood risk estimation can be easily produced by Geographic Information System (GIS) or extracted from remote sensing images, with most of these indicators adopting pixels as the basic estimation unit (Bathrellos et al., 2017; Huu et al., 2020; Kanani-Sadat et al., 2019; Skilodimou et al., 2019). In AHP, the definition of the judgment matrix determines the final weight of the chosen estimation indicators (Saaty, 2004). Therefore, even a tiny difference among criteria weight definition can lead to a considerable impact on the estimation results (Ohnishi et al., 2011; Ohnishi and Imai, 1998). Thus, the judgment matrix

2
Submmited to pre-review

becomes dominant in flood risk estimation (Adar et al., 2021; Zarei et al., 2021). Efforts are still needed to reduce the impact of criteria weight sensitivity on the risk estimation results.

In flood risk estimation, when adopting sub-watershed as a basic unit, the differences among individual cells inside sub-watersheds are considered as coincident areas. Thus, using sub-watershed as a basic unit can, to a certain degree, reduce the flood risk estimation sensitivity from disparity in criteria weight definition. In the flood simulation domain, sub-watershed is a widely used boundary condition for rainfall-runoff simulation (Abdulkareem et al., 2018; Shao et al., 2019; Wang et al., 2020). During natural flood processes, the flood risk of cells is more dependent on the maximum risk level of neighborhood cells at a sub-watershed scale rather than the terrain features or hydrological characteristics of individual cells (Betancourt-Suárez et al., 2021; Majeed et al., 2021; Noaa, 2019). In our previous effort, we proposed a watershed-based AHP model, termed WZSAHP, which used sub-watershed as the basic unit to constraint runoff converging related indicators (Zhang et al., 2021). The results indicated that WZSAHP outperformed other pixel-based AHP under the same judgment matrix definition.

In this study, we, therefore, further investigate the sensitivity of flood risk estimation given the disparity in the weight criteria using sub-watershed-based WZSAHP and pixel-based AHP. We compared the flood risk estimation sensitivity of pixel-based AHP model to four types of WZSAHP models: MFD-RC, MFD-All, D8-RC, and D8-All. Detailed definitions of these models are given in Section 3. Taking a flood event that occurred in July 2020 in Chaohu basin, Anhui province, China, as a study case, we used the flood areas extracted from remote sensing images to construct ground truth for validation purposes, aiming to explore the sensitivity of flood risk estimation caused by criteria definition. This paper is organized as follows. Section 2 introduces the study area and the main data sources. Section 3 describes the methodology of our study. Section 4 presents the results and discusses the major findings. Sections 5 and 6 present the limitations of this study and conclusions.

## 2. Study area and materials

### 2.1. Study area

Lying on the downstream of the Yangtze River, the Chaohu basin in Anhui province, China, suffers a high probability of floods during the rain-rich seasons between June to August every year. The Chaohu basin contains several rivers that are connected to Chao Lake, such as the Hangfu river, Fengle river, Zhao river, Xi river, Nanfei river, Pai river. During the Meiyu (June to August) period in 2020, the Chaohu basin reached the highest water level in history (13.48m) on July 22, 2020, due to continuous and intense rainfall, posing great threats to the cities of Chaohu, Hefei. In this study, we take the Chaohu basin as a study case to analyze flood risk estimation sensitivity to the judgment matrix, aiming to improve the flood risk estimation accuracy and mitigate flood-related losses. The study area is defined as the region that surrounds Chao Lake with a 10-kilometer buffer, as shown in Fig.1.

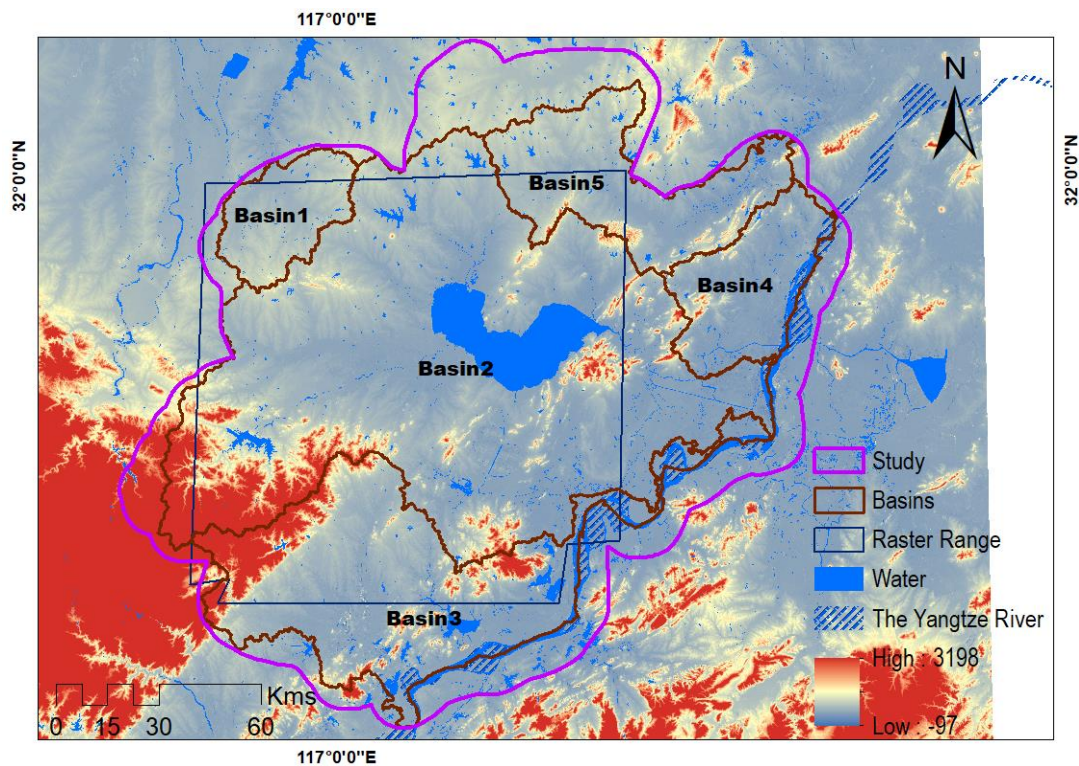

Figure 1. The geographical location of the study area (according to the China basic geographic information, 2008 version).

Fig.1 presents the geographical location of the study area and the Digital Elevation Model (DEM). The flooded areas were extracted from remote sensing images in July 2020. There exist five main basins in the study area, with the *Basin1* and *Bain5* flowing out from the study area, while the *Basin2*, *Basin3*, and *Basin4* flowing into the Yangtze River. Among these five basins, *Basin2* that contains Chao Lake has the largest size. The Nanfei River, the Fengle River, and the Hangfu River, etc., all converge to Chao Lake and then flow through the Yuxi river to the Yangtze River. The Xi River in *Basin2* also meets the Yuxi River and eventually flows to the Yangtze River.

*2.2. Materials*

Table 1 presents the major data sources and their descriptions. We used geographic information system (GIS) vectors to present hydrological information. A DEM dataset was used to divide watersheds and calculate slopes. Impervious surface products were used to extract land cover and hydrological infiltration information. SAR (synthetic aperture radar) images were used to extract flooded areas, serving as the grounth truth. All the pre-processing steps that include transforming, projecting, mosaicking, and clipping were implemented in ArcGIS 10.3. We also used flood-related Baidu News to validate flood events that occurred at the town scale.

Table 1. The datasets used in this study.

| Data Sources | Used Data | Detailed Information |
|---|---|---|
| Geographic information (1:1million) | District, hydrological layers | We used the vectorized distinct boundaries and hydrological layers released in 2008. When delimitating sub-watersheds, the Hydrological layers, including rivers, streams, and lakes, were used to constraint terrain as natural water bodies. |
| ASTER GDEM V2 (30m) | DEM | The DEM was downloaded from *http://www.gscloud.cn*. The DEM is the main material used to divide watersheds and extract digital streams. It is also used to derive flood risk estimation indicators, such as the slope and the elevation. |
| China's impervious surface product (2m) | water, vegetation, soil, building, and roads | We adopted China's impervious surface grid product (2m) (Shao et al., 2019) that contains classification types of water, vegetation, soil, building, and roads. Land use types that include water, impervious surface, and pervious surfaces were used to prepare hydrological indicators. |
| Images for extracting flooded areas | Water bodies | We used Landsat 8 OLI on July 20, 2020, and GF-3 on July 24, 2020, to extract the flooding areas. The Landsat 8 OLI was downloaded from [https://www.usgs.gov](https://www.usgs.gov). The GF-3 data is supported by the GaoFen center of Hubei province. The flooded areas were used to prepare ground truth for validation purposes. Nature water bodies were excluded from flooded areas. |
| Flood-relevant information in Baidu News | Flooding events and dam breaks by towns | We used an internet context searching and capturing tool named "Octopus" [(https://www.bazhuayu.com)](https://www.bazhuayu.com) to collect information on floods and dikes during the flood event in the Chaohu basin that occurred in July 2020 from the Baidu News website [(https://news.baidu.com)](https://news.baidu.com). |

## *3. Methodology*

In this paper, we evaluated the sensitivity of flood risk estimation sensitivity given different judgment definitions by comparing pixel-based AHP (pixel-AHP) and sub-watershed-based WZSAHP. The sub-watershed-based WZSAHP method was adopted from our previous work (Zhang et al., 2021). We defined a total of 48 judgment matrixes and used them to estimate flood risk levels via the pixel-AHP and the WZSAHP models, including MFD-RC, MFD-All, D8-RC, and D8-All.

The MFD-RC model uses a multiple flow direction algorithm (MFD) (Zhang et al., 2020) to delimitate sub-watersheds for constraining converging related indicators, while the MFD-All adopts MFD to divide sub-watershed to constraint all indicators. The D8-RC model uses sub-watersheds derived by the D8 algorithm to constraint part of indicators, while the D8-All model constraints all indicators via D8 delimitated sub-watersheds. We evaluated the accuracy among these models using ground-truthing flooded areas extracted from remote sensing imagery and analyzed the influences of different judgment definitions on model accuracy. The technical workflow of this study is illustrated in Fig.2.

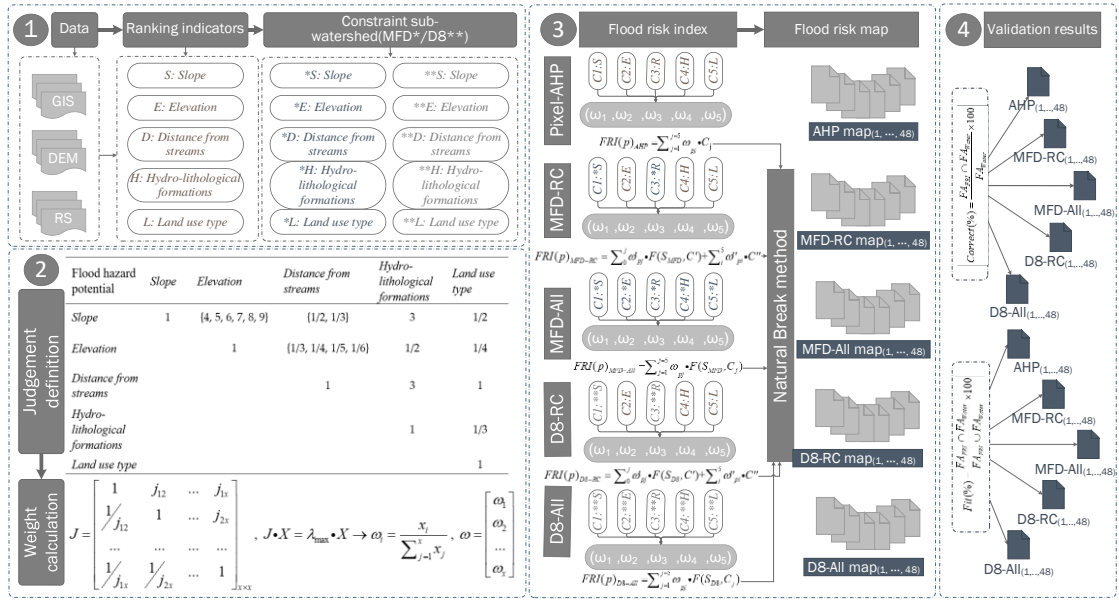

Figure 2. The overall workflow of deriving flood risk maps and validation datasets in our study.

Fig.2 shows the main process of this study that contains four major parts (labeled as numbers 1-4 in the workflow). Part one represents the preparation process of flood risk estimation indicators. In this study, we defined the *Slope*, *Elevation*, *Distance from streams*, *hydro-lithological formations*, and *Land used types* as indicators. These indicators were used in their original ranked form or constrained by sub-watersheds delimitated via MFD and D8 methods. Part two represents the construction of the judgment matrix and the calculation of weights. Referring to (Zhang et al., 2021), the importance of these indicators follows *Slope > Distance from streams > Elevation*. The responding weight vectors can be calculated according to judgment matrixes. Part three represents the process of flood risk estimation that includes flood risk index calculation and flood risk mapping. A total of five methods, i.e., AHP, MFD-RC, MFD-All, D8-RC, and D8-All, were used to calculate the flood risk index. The results of calculated flood risk indexes were sliced via the Natural Breaking method. Part four represents the result validation process. The correct ratio and fit ratio were used to validate the accuracy of estimated results by comparing them with the ground-truthing flooded areas.

### 3.1. The basic theory of used AHP models

In this study, pixel-based AHP model and the sub-watershed-based WZSAHP models (Zhang et al., 2021) were used to derive flood risk levels. The pixel-AHP adopts pixels as the basic unit and constructs flood risk estimation assisted by pixel-scale weight vectors from a group of raster layers. Flood risk estimation via pixel-AHP and sub-watershed-based WZSAHP have the process: choosing indicators of flood risk estimation → preparing raster layers of indicators → defining judgment matrix and deriving weight vectors → deriving flood risk maps. However, different from the pixel-AHP, the WZSAHP constrains relevant indicators using sub-watersheds.

- *The pixel based AHP model*

AHP is composed of three levels, i.e., target, criteria, and alternatives, as shown in Fig.3.

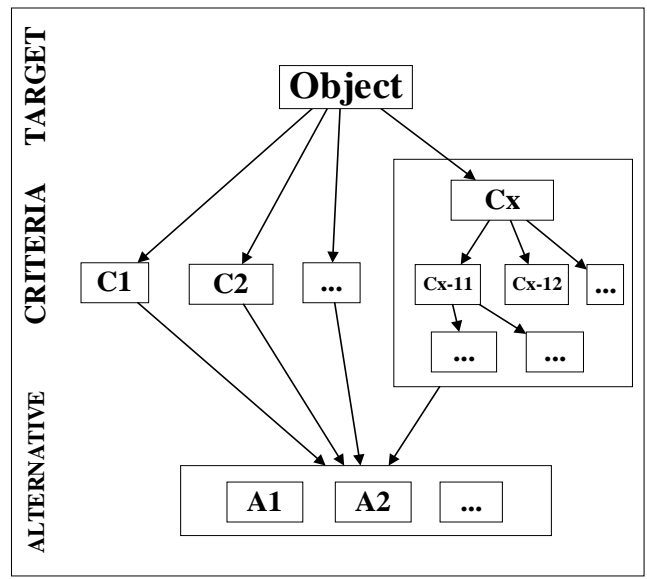

Figure 3. The structure of the AHP method.

As shown in Fig.3, the target layer refers to the evaluation unit; the criteria (with single or multiple layers) consist of several clusters that reflect different aspects of the target; the alternative is composed of the estimated results. The flood risk estimation target, the criteria and the corresponding watershed are defined as follows:

$$P = \begin{bmatrix} p_{11} & p_{12} & \cdots & p_{1n} \\ p_{21} & p_{22} & \cdots & p_{2n} \\ \cdots & \cdots & \cdots & \cdots \\ p_{m1} & p_{m2} & \cdots & p_{mn} \end{bmatrix}, C = \begin{bmatrix} c_1 \\ c_2 \\ \cdots \\ c_x \end{bmatrix}, C_x = \begin{bmatrix} c_{11} & c_{12} & \cdots & c_{1n} \\ c_{21} & c_{22} & \cdots & c_{2n} \\ \cdots & \cdots & \cdots & \cdots \\ c_{m1} & c_{m2} & \cdots & c_{mn} \end{bmatrix}_x \quad (1)$$

where matrix $P$ represents the pixels in the study area, with a size of $m \times n$, $C$ is the flood risk estimation indicators, with each of the indicator $C_x$ suggesting a raster layer with a size of $m \times n$.

A comparative matrix of criteria and calculated their weights are further constructed. According to the AHP model, the positive pairwise comparison matrix uses values from 1 to 9 to indicate the relative importance of two indices (as shown in Table 2), and its eigenvector that corresponds to the largest eigenvalue can be used as a weight vector to represent the established hierarchic evaluation structure (Saaty, 2004). The hierarchic evaluation structure can be formulated as follows:

$$J = \begin{bmatrix} 1 & j_{12} & \cdots & j_{1x} \\ 1/j_{12} & 1 & \cdots & j_{2x} \\ \cdots & \cdots & \cdots & \cdots \\ 1/j_{1x} & 1/j_{2x} & \cdots & 1 \end{bmatrix}_{x \times x}, \quad J \bullet X = \lambda_{\max} \bullet X \rightarrow \omega_i = \frac{x_i}{\sum_{j=1}^{x} x_j}, \quad \omega = \begin{bmatrix} \omega_1 \\ \omega_2 \\ \cdots \\ \omega_x \end{bmatrix} \quad (2)$$

where the comparison matrix $J$, with a size of $x \times x$, is used to determine the importance order among criteria $C$ (in Eq. (1)). $X$ is the eigenvector corresponding to the largest eigenvalue $\lambda_{\max}$ of $J$, $\omega$ is the weight vector corresponding to the normalized value of eigenvector $X$.

Table 2. The scales for pairwise comparison (Referring to (Saaty, 2004))

| Important levels | Description | Explanation |
| --- | --- | --- |
| 1 | Equal importance | Two activities contribute equally to the objective |
| 3 | Moderate importance | Experience and judgment slightly favor one activity over another. |
| 5 | Strong importance | Experience and judgment strongly favor one activity over another. |
| 7 | Very strong importance | An activity is favored very strongly over another; its dominance demonstrated in practice. |

| 9 | Extreme importance | The evidence favoring one activity over another is of the highest possible order of affirmation. |
| 2, 4, 6, 8 | Intermediate values | The intermediate values of the above values. |
| Reciprocals | Inverse comparison | A reasonable assumption. |

To keep the order of relative importance among criteria logically consistent, the consistency ratio ($CR$) can be calculated as in Eq. (3). The consistency ratio of a pairwise comparison matrix is the ratio of its consistency index to the corresponding random index value in Table.3 (Saaty, 2004). The pairwise comparison matrix can be accepted if its consistency ratio is less than 0.1 (a consistency ratio of 0 indicates that the judgment matrix is completely consistent).

$$CI = \frac{\lambda_{max} - n}{n - 1}, \quad CR = \frac{CI}{RI} \quad (3)$$

where $CR$ is the consistency ratio, $CI$ is the consistency index, $RI$ is a statistic random index, $RI$ is the average $CI$ of randomly generated pairwise of comparison matrix of similar size (as shown in Table 3), $\lambda_{max}$ is the largest eigenvalue of the comparison matrix, $n$ is the number of indicators used in criteria.

Table 3. Random index (Refering to (Saaty, 2004)).

| n | 1 | 2 | 3 | 4 | 5 | 6 | 7 | 8 | 9 | 10 |
|---|---|---|---|---|---|---|---|---|---|---|
| Random Index | 0 | 0 | .52 | .89 | 1.11 | 1.25 | 1.35 | 1.40 | 1.45 | 1.49 |

- ***The sub-watershed-based WZSAHP model***

Referring to (Zhang et al., 2021), the WZSAHP flood risk estimation method adopts sub-watersheds as the constraint unit to calculate the maximum zonal statistical value of relevant indicators. The runoff converging related indices, such as *slope* and *distance from a stream*, are constrained by sub-watersheds via Eq.(4)-(5), and the used flood risk index can be calculated via Eq. (6). Thus, the final flood risk distribution can be mapped according to the flood risk index sliced by the Nature Break method, as {"*Very Low*", "*Low*", "*Normal*", "*High*" and "*Very High*"}:

$$S = \begin{bmatrix} \cdots & & & \\ & S_k & S_k & \\ & & S_k & \\ & & & \cdots \end{bmatrix}_{m \times n} \quad (4)$$

$$F(S, c_x) = zonalStatistic(S, c_x, Maximum) \quad (5)$$

where $S$ is the sub-watershed division raster, $F(S, c_x)$ is constraint sub-watershed as a statistical zonal unit with updated corresponding indicator $c_x$. Note that the size of $F(S, c_x)$ is also $m \times n$. $zonalStatistic$ is calculated using the descriptive statistics of indicator $c_x$ for each sub-watershed $S$, $Maximum$ suggests the maximum statistical method.

$$FRI = \omega \bullet C = \sum_{i=1}^{i=m} \omega_i \bullet F(S, c_i) + \sum_{j=m+1}^{j=x} \omega_j \bullet c_j, \quad (0 \leq m \leq x) \quad (6)$$

where $FRI$ is the flood risk index calculated by the cumulative sum of criteria $C$ (using either original indices $c_j$ or sub-watershed constraint indices $F(S, c_i)$) and its corresponding weight $\omega$.

### *3.2. Choosing indicators for flood risk estimation*

In flood risk estimation, studies have shown that information regarding topography, hydrology, and geological location are considerably dominant(Kazakis et al., 2015; Nachappa et al., 2020; Pham et al., 2021; Shariat et al., 2019). For example, the terrain features that include the slope distribution and the local distances from certain sources are the main factors that drive the distribution patterns of flood risks. The terrain features, such as elevation, slope, hydrological systems (e.g., lakes, reservoirs, rivers, and the low-lying wetland areas), may determine rainwater converging path, thus influencing situ likelihood of occurring floods. Furthermore, the rainfall-runoff production-related factors (e.g., land use type, porous and impervious features) drive the rainwater carrying ability, and the flow-out and effective drainage abilities, also impacting flood risk distributions.

In this study, we adopted five indices to construct flood risk estimation criteria $C = \{C_1, C_2, C_3, C_4, C_5\}$, where $C_1$= *Slope*, $C_2$= *Elevation*, $C_3$= *Distance from streams*, $C_4$= *Hydro-lithological formations*, $C_5$= *Land use type*, following (Bathrellos et al., 2017).

- *Slope*. The slope is the main factor that influences the rainwater flow path.
- *Elevation*. The elevation also influences flood risk distribution.
- *Distance from streams*. Streams are the source of flood risks. The distance from streams reveals the potential flood risk.
- *Hydro-lithological formations*. Hydro-lithological formations determine infiltration capability in rainfall-runoff production.
- *Land use types*. The land use types determine the rainfall-runoff production.

### 3.3. Preparing raster layers of indicators

The original indicators collected from different sources with varying numeric distributions described different aspects of flood risks. The flood risk index in this study is calculated by accumulating the value after multiplying indicators with their corresponding weight. Thus, ranking the original indicators is a necessary pre-processing in preparing flood risk indicators. Here, all the flood risk estimation indicators are ranked, as shown in Table 4.

Table 4. The classes and rating of involved.

| Factors | Classes | Rating | Factors | Classes | Rating |
|---|---|---|---|---|---|
| *Slope( °)* | 0 | 5 | *Hydro lithological formations* | Water | 4 |
| | 0 – 2 | 4 | | Impervious surface | 3 |
| | 2 – 6 | 3 | | Pervious surface | 1 |
| | 6 – 12 | 2 | | | |
| | 12 - 20 | 1 | | | |
| | >20 | 0 | | | |
| *Elevation(m)* | - 204 - 12 | 5 | *Land use types* | Water | 5 |
| | 12 - 23 | 4 | | Road | 4 |
| | 23 - 46 | 3 | | Building | 3 |
| | 46 - 152 | 2 | | Soil | 2 |
| | >152 | 1 | | Vegetation | 1 |

| Factors | Classes | | | | | Rating |
|---|---|---|---|---|---|---|
| Distance from streams(m) | Rivers, lakes and reservoirs | | | | | 5 |
| | Level 1 | Level 2 | Level 3 | Level 4 | Level 5 | |
| | | | | | 0 – 1,000 | 4 |
| | | | 0 - 500 | 0 – 1,000 | 1,000 – 2,000 | 3 |
| | | 0 - 500 | 500 – 1,000 | 1,000 – 2,000 | 2,000 – 4,000 | 2 |
| | 0 - 500 | 500 – 1,000 | 1,000 – 1,500 | 2,000 – 3,000 | 4,000 – 6,000 | 1 |
| | >500 | >1,000 | >1,500 | >3000 | >6,000 | 0 |

As shown in Table 4, ranked by the Natural Breaking method, *Slope* and *Elevation* are numeric values and distributed in a nearly normal fashion. For the factors *Distance from streams*, *Land use types*, and *Hydro lithological formations*, we ranked them according to expert experiences. The distribution patterns of ranked indicators are shown in Fig.4.

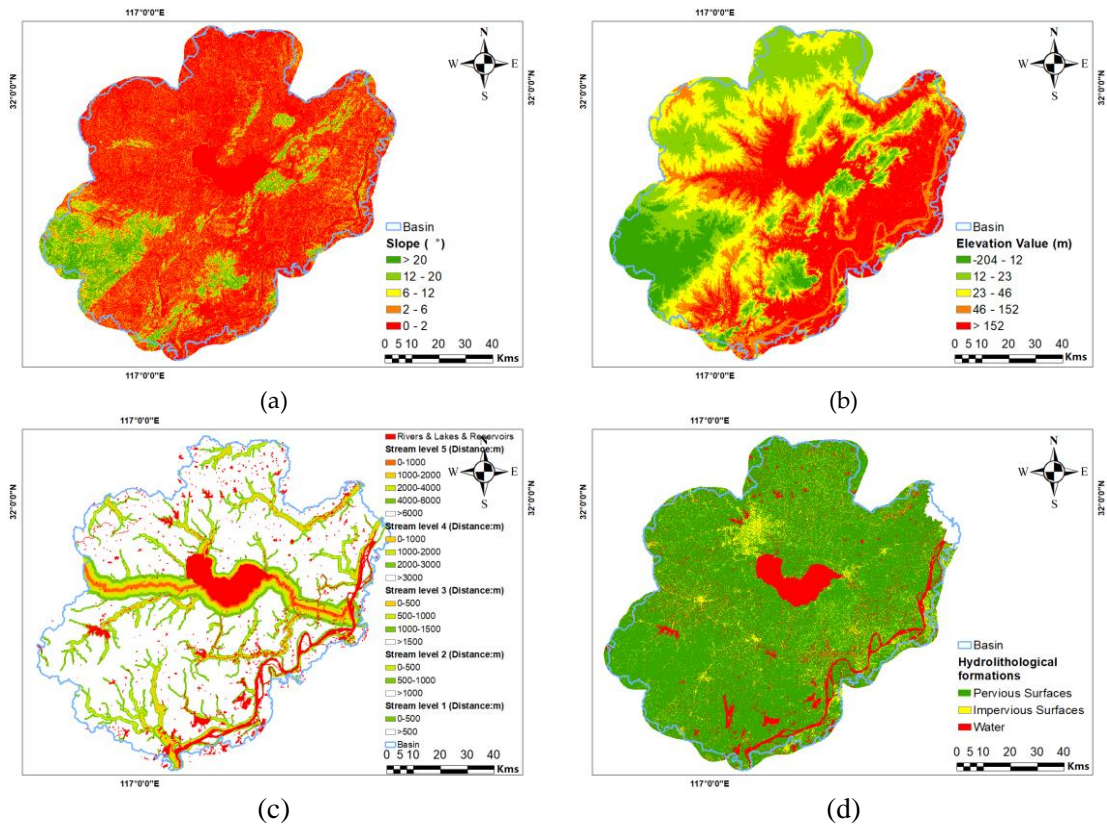

(a)　　　　　　　　　　　　(b)

(c)　　　　　　　　　　　　(d)

Fig.4. Cont.

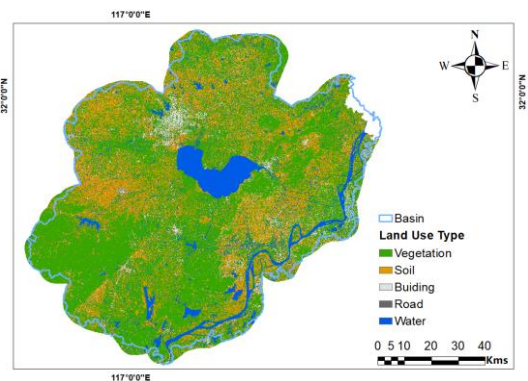

(e)

Figure 4. The spatial distribution of ranked (re-classified) factors involved in the flood risk analysis: (a)Slope, (b) Elevation, (c) Distance from streams, (d) Land use type, (e) Hydro lithological formations.

Fig. 4(a) shows the *Slope* factor. The original range of slope is 0 °- 81 °, and it is classified into six types by the Natural Break method, the angles of "0", "0 - 2", "2 - 6", "6 - 12", "12 - 20", and "> 20" were labeled as 5, 4, 3, 2, 1 and 0, respectively. Fig. 4(b) shows the *Elevation* factor that ranges from -204 to 1,490 m. *Elevation* was grouped into five types via the Natural Break method. The elevation of "-204 - 12", "12 - 23", "23 - 46", "46 - 152", and ">152" were labeled as 5, 4, 3, 2 and 1, respectively. Fig. 4(c) shows the *Distance from stream*s distribution. The streams were derived by watershed delimitation function using the hydrological tool in ArcGIS 10.3. The distances involved in the reclassification of *Distance from streams* can be found in Table 4. Fig. 4(d) shows the *Hydro-lithological formations* distribution. We ranked the water, impervious surface and pervious surface as 4, 3, and 1, respectively. Fig. 4(e) shows the land use type factor. The impervious surfaces extracted by GF-2 remote sensing (Shao et al., 2019)(Shao et al., 2019)(Shao et al., 2019)(Shao et al., 2019) were used in this study. We ranked vegetation, soil, building, road, and water as 1, 2, 3, 4 and 5, respectively.

For WZSAHP, an addition preparing step was implemented, i.e., using sub-watershed to calculate the statistical maximum value of related ranked indicators. In this study, the MFD-All model adopts MFD delimitated sub-watershed to constraint the indicators, while the MFD-RC model only constraints converging related indicators (i.e., *Slope* and *Distance from Streams*). The D8-All model constrains all indicators by D8 delimitated sub-watershed, while the D8-RC model

Submmited to pre-review

constrains only converging related indicators (i.e., *Slope* and *Distance from Streams*). Taking the MFD-derived sub-watershed and D8-derived sub-watershed as the basic unit, we calculated the maximum statistical value of the original ranked indicators. This process was conducted via the zonal statistical function of the spatial analyses tool in ArcGIS 10.3.

### *3.4. Definition pairwise comparison elements in judgment matrix*

The geological indicators that include *Slope*, *Elevation*, and *Distance from streams* describe rainwater runoff convergence path characteristics, while *Hydro-lithological formations* and *Land use type* reflect the rainfall-runoff ratio.

In this study, we focus on investigating the sensitivity in pairwise comparison targeting flood converging related factors. That is to say, the couples of pairwise comparison elements among *Slope*, *Elevation*, and *Distance from streams* are changing while others remain the same. Here, we design projects to reflect the pairwise elements of (*Slope*, *Elevation*), (*Slope*, *Distance from streams*), and (*Elevation*, *Distance from streams*). In a study by Zhang et al. (2021), these three indicators are ordered following *Distance from streams > slope > Elevation*. Table 5 presents the candidate judgment matrix.

Table 5. The judgment matrix of criteria.

| Flood hazard potential | Slope | Elevation | Distance from streams | Hydro-lithological formations | Land use type |
|---|---|---|---|---|---|
| *Slope* | 1 | {4, 5, 6, 7, 8, 9} | {1/2, 1/3} | 3 | 1/2 |
| *Elevation* | | 1 | {1/3, 1/4, 1/5, 1/6} | 1/2 | 1/4 |
| *Distance from streams* | | | 1 | 3 | 1 |
| *Hydro-lithological formations* | | | | 1 | 1/3 |
| *Land use type* | | | | | 1 |

As shown in Table 5, the constant couple of pair-wise comparison elements values between {*Slope*} and {*Hydro-lithological formation*, *Land use type*} are {3, 1/2}, between {*Elevation*} and {*Hydro-lithological formation*, *Land use type*} are {1/2, 1/4}, between {*Distance from streams*} and {*Hydro-lithological formation*, *Land use type*} are {3, 1}, and pair-wise element of (*hydro-lithological formation*, *land use type*) is {1/3}. Meanwhile, the variable pairwise comparison elements were: 1) six types of pairwise element (*Slope*, *Elevation*) in a set of {4, 5, 6, 7, 8, 9}, 2) two types of pairwise element of (*Slope*, *Elevation*) in a set of {1/2, 1/3}, and 3) four types of pairwise element (*Slope*, *Elevation*) in a set of {1/3, 1/4, 1/5, 1/6}. Therefore, we have a total of 48 projects (i.e., 48 different matrix definitions), calculated as $48 = C_6^1 \cdot C_2^1 \cdot C_4^1$. The final pairwise elements of candidate judgment matrix definitions and their responding weight vectors are illustrated in Table 6.

Table 6. Definitions of pairwise elements in the judgment matrix and the calculated weight vector, maximum eigenvalue, and consistency ratio.

| Prj. No. | S/E | S/R | E/R | ω1 | ω2 | ω3 | ω4 | ω5 | $\lambda_{max}$ | CI |
|---|---|---|---|---|---|---|---|---|---|---|
| 1 | 4 | 1/2 | 1/3 | 0.214 | 0.068 | 0.302 | 0.1 | 0.315 | 5.133 | 0.030 |
| 2 | 4 | 1/2 | 1/4 | 0.211 | 0.063 | 0.314 | 0.099 | 0.314 | 5.097 | 0.022 |
| 3 | 4 | 1/2 | 1/5 | 0.208 | 0.06 | 0.323 | 0.098 | 0.311 | 5.084 | 0.019 |
| 4 | 4 | 1/2 | 1/6 | 0.206 | 0.057 | 0.331 | 0.097 | 0.309 | 5.081 | 0.018 |
| 5 | 4 | 1/3 | 1/3 | 0.197 | 0.068 | 0.331 | 0.098 | 0.306 | 5.212 | 0.047 |
| 6 | 4 | 1/3 | 1/4 | 0.194 | 0.062 | 0.342 | 0.097 | 0.305 | 5.165 | 0.037 |
| 7 | 4 | 1/3 | 1/5 | 0.191 | 0.058 | 0.351 | 0.096 | 0.303 | 5.143 | 0.032 |
| 8 | 4 | 1/3 | 1/6 | 0.188 | 0.056 | 0.359 | 0.095 | 0.301 | 5.134 | 0.030 |
| 9 | 5 | 1/2 | 1/3 | 0.223 | 0.065 | 0.3 | 0.098 | 0.313 | 5.163 | 0.036 |
| 10 | 5 | 1/2 | 1/4 | 0.22 | 0.06 | 0.311 | 0.097 | 0.311 | 5.121 | 0.027 |
| 11 | 5 | 1/2 | 1/5 | 0.217 | 0.057 | 0.321 | 0.096 | 0.31 | 5.104 | 0.023 |
| 12 | 5 | 1/2 | 1/6 | 0.214 | 0.054 | 0.329 | 0.096 | 0.308 | 5.098 | 0.022 |
| 13 | 5 | 1/3 | 1/3 | 0.206 | 0.064 | 0.33 | 0.096 | 0.303 | 5.251 | 0.056 |
| 14 | 5 | 1/3 | 1/4 | 0.202 | 0.059 | 0.341 | 0.095 | 0.302 | 5.197 | 0.044 |
| 15 | 5 | 1/3 | 1/5 | 0.199 | 0.056 | 0.35 | 0.095 | 0.301 | 5.171 | 0.038 |
| 16 | 5 | 1/3 | 1/6 | 0.196 | 0.053 | 0.358 | 0.094 | 0.299 | 5.158 | 0.035 |
| 17 | 6 | 1/2 | 1/3 | 0.232 | 0.063 | 0.298 | 0.097 | 0.31 | 5.198 | 0.044 |
| 18 | 6 | 1/2 | 1/4 | 0.228 | 0.058 | 0.309 | 0.096 | 0.309 | 5.151 | 0.034 |
| 19 | 6 | 1/2 | 1/5 | 0.224 | 0.055 | 0.318 | 0.095 | 0.308 | 5.129 | 0.029 |
| 20 | 6 | 1/2 | 1/6 | 0.221 | 0.052 | 0.326 | 0.094 | 0.306 | 5.121 | 0.027 |
| 21 | 6 | 1/3 | 1/3 | 0.214 | 0.062 | 0.329 | 0.095 | 0.3 | 5.293 | 0.065 |
| 22 | 6 | 1/3 | 1/4 | 0.21 | 0.057 | 0.34 | 0.094 | 0.3 | 5.233 | 0.052 |
| 23 | 6 | 1/3 | 1/5 | 0.206 | 0.054 | 0.348 | 0.093 | 0.299 | 5.203 | 0.045 |
| 24 | 6 | 1/3 | 1/6 | 0.203 | 0.051 | 0.356 | 0.093 | 0.297 | 5.188 | 0.042 |
| 25 | 7 | 1/2 | 1/3 | 0.239 | 0.061 | 0.296 | 0.095 | 0.308 | 5.234 | 0.052 |
| 26 | 7 | 1/2 | 1/4 | 0.235 | 0.056 | 0.307 | 0.094 | 0.307 | 5.183 | 0.041 |
| 27 | 7 | 1/2 | 1/5 | 0.231 | 0.053 | 0.316 | 0.094 | 0.306 | 5.158 | 0.035 |
| 28 | 7 | 1/2 | 1/6 | 0.228 | 0.05 | 0.324 | 0.093 | 0.304 | 5.147 | 0.033 |
| 29 | 7 | 1/3 | 1/3 | 0.221 | 0.06 | 0.328 | 0.093 | 0.298 | 5.336 | 0.075 |
| 30 | 7 | 1/3 | 1/4 | 0.217 | 0.055 | 0.338 | 0.092 | 0.297 | 5.272 | 0.061 |
| 31 | 7 | 1/3 | 1/5 | 0.213 | 0.052 | 0.347 | 0.092 | 0.296 | 5.238 | 0.053 |
| 32 | 7 | 1/3 | 1/6 | 0.21 | 0.049 | 0.354 | 0.091 | 0.295 | 5.220 | 0.049 |
| 33 | 8 | 1/2 | 1/3 | 0.247 | 0.059 | 0.295 | 0.094 | 0.306 | 5.271 | 0.061 |
| 34 | 8 | 1/2 | 1/4 | 0.242 | 0.054 | 0.305 | 0.093 | 0.305 | 5.216 | 0.048 |
| 35 | 8 | 1/2 | 1/5 | 0.238 | 0.051 | 0.314 | 0.093 | 0.304 | 5.188 | 0.042 |
| 36 | 8 | 1/2 | 1/6 | 0.235 | 0.049 | 0.322 | 0.092 | 0.303 | 5.175 | 0.039 |
| 37 | 8 | 1/3 | 1/3 | 0.228 | 0.059 | 0.327 | 0.092 | 0.295 | 5.380 | 0.085 |
| 38 | 8 | 1/3 | 1/4 | 0.223 | 0.054 | 0.337 | 0.091 | 0.295 | 5.311 | 0.069 |
| 39 | 8 | 1/3 | 1/5 | 0.219 | 0.05 | 0.345 | 0.09 | 0.294 | 5.273 | 0.061 |
| 40 | 8 | 1/3 | 1/6 | 0.216 | 0.048 | 0.353 | 0.09 | 0.293 | 5.253 | 0.056 |
| 41 | 9 | 1/2 | 1/3 | 0.254 | 0.058 | 0.293 | 0.093 | 0.304 | 5.308 | 0.069 |
| 42 | 9 | 1/2 | 1/4 | 0.249 | 0.053 | 0.303 | 0.092 | 0.303 | 5.249 | 0.056 |
| 43 | 9 | 1/2 | 1/5 | 0.244 | 0.05 | 0.312 | 0.092 | 0.302 | 5.219 | 0.049 |
| 44 | 9 | 1/2 | 1/6 | 0.241 | 0.048 | 0.319 | 0.091 | 0.301 | 5.204 | 0.046 |
| 45 | 9 | 1/3 | 1/3 | 0.235 | 0.057 | 0.325 | 0.09 | 0.293 | 5.423 | 0.095 |
| 46 | 9 | 1/3 | 1/4 | 0.23 | 0.052 | 0.336 | 0.09 | 0.293 | 5.350 | 0.078 |
| 47 | 9 | 1/3 | 1/5 | 0.225 | 0.049 | 0.344 | 0.089 | 0.292 | 5.309 | 0.069 |
| 48 | 9 | 1/3 | 1/6 | 0.222 | 0.047 | 0.351 | 0.089 | 0.291 | 5.286 | 0.064 |

As shown in Table 6, the columns *S/E, S/R,* and *E/R* represent the pairwise elements of (*Slope, Elevation*), (*Slope, Distance from streams*), and (*Elevation, Stream*), respectively. The weight vectors from the 48 projects are shown in Fig.6. The detailed definitions of judgment matrixes and related information can be found in appendix B (Fig. B.1).

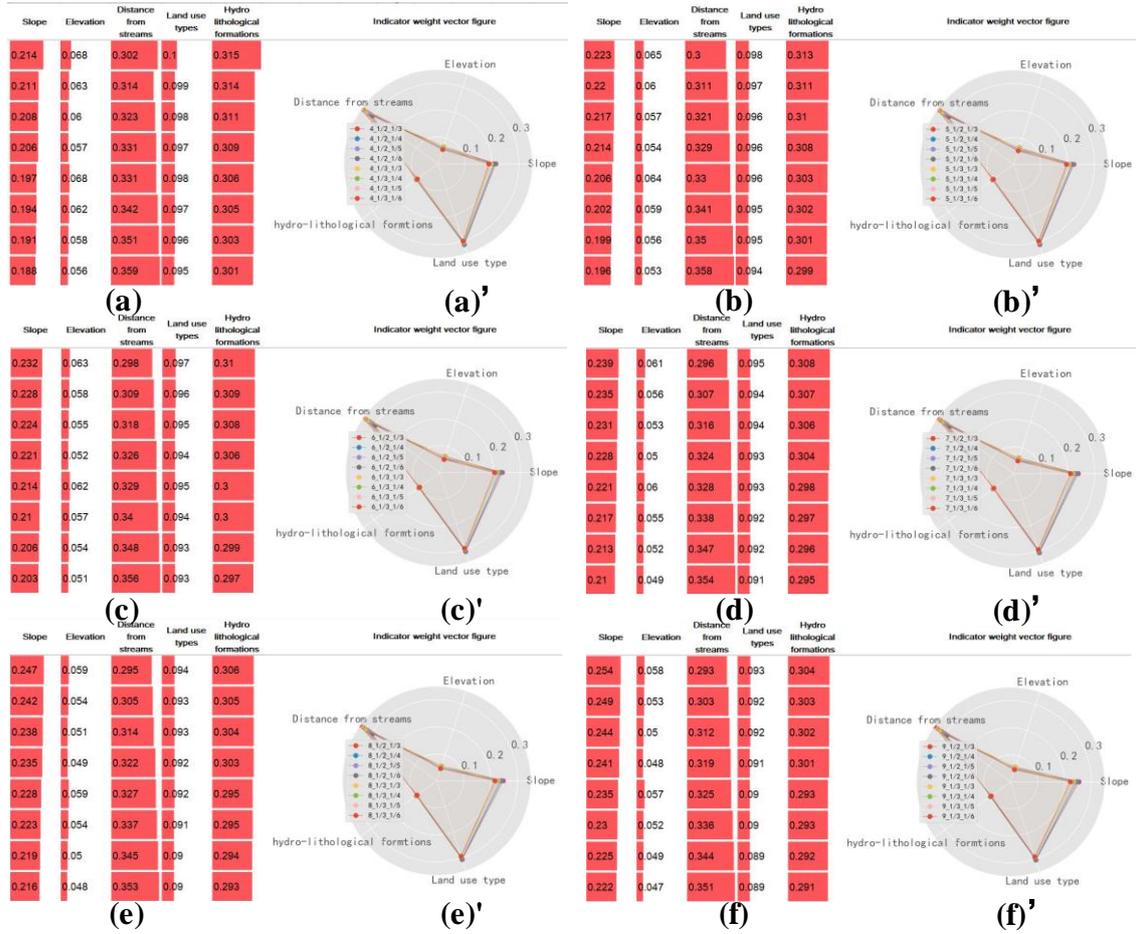

Figure 5. The weight vectors derived by custom judgment matrixes and the related radar maps grouped by the value of pairwise comparison element of (*Slope, Elevation*). Figures (a)-(f) present the weight vectors, while figures (a)'-(f)' present the radar maps for each group of weight vectors.

As shown in Fig.5, weight vectors differ with different judgment matrixes. Two major orders can be found, i.e., (*Distance from streams > Slope > Elevation*) and (*Hydro-lithological formation > Land use type*). Among each cluster definition of (*Slope*, *Elevation*), the inner differences show that the weight of *Slope* decreases with the increase of weight of *Distance from streams*, as the definition of ((*Slope*, *Distance from streams*), (*Elevation*, *Distance from streams*)) changes in a set of {(1/2, 1/3), (1/2, 1/4), (1/2, 1/5), (1/2, 1/6), (1/3, 1/3), (1/3, 1/4), (1/3, 1/5), (1/3, 1/6)}.

### 3.5. *Deriving flood risk maps*

After preparing flood estimation indicators and calculating the weight vectors of the candidate judgment matrix, the flood risk index among the five models can be calculated. For a total of 48 projects of pixel-AHP, MFD-RC, MFD-All, D8-RC, and D8-All models, their flood risk indexes were calculated by Eq. (7)-(11).

$$FRI(p)_{AHP} = \sum_{j=1}^{j=5} \omega_{pj} \cdot C_j, \; p \in (1,48) \tag{7}$$

$$FRI(p)_{MFD-All} = \sum_{j=1}^{j=5} \omega_{pj} \cdot F(S_{MFD}, C_j), \; p \in (1,48) \tag{8}$$

$$FRI(p)_{MFD-RC} = \sum_{0}^{j} \omega'_{pj} \cdot F(S_{MFD}, C') + \sum_{i}^{5} \omega''_{pi} \cdot C'', \; p \in (1,48) \tag{9}$$

$$FRI(p)_{D8-All} = \sum_{j=1}^{j=5} \omega_{pj} \cdot F(S_{D8}, C_j), \; p \in (1,48) \tag{10}$$

$$FRI(p)_{D8-RC} = \sum_{0}^{j} \omega'_{pj} \cdot F(S_{D8}, C') + \sum_{i}^{5} \omega''_{pi} \cdot C'', \; p \in (1,48) \tag{11}$$

where, given the project number $p$, the $FRI(p)_{AHP}$ is the flood risk index used for pixel-AHP model. The $FRI(p)_{MFD-RC}$, $FRI(p)_{MFD-All}$, $FRI(p)_{D8-RC}$, and $FRI(p)_{D8-All}$ are the flood risk indexes of MFD-RC, MFD-All, D8-RC, and D8-All models, respectively. The criteria are marked as $C = \{C_1, C_2, C_3, C_4, C_5\}$. $\omega_{pj}$ is the weight of indicator $C_j$ of project number $p$. Information on $F(S,C)$ can be found in Eq.(5). $S_{D8}$ represents sub-watershed delimitated by D8 while $S_{MFD}$ represents sub-watershed delimitated by MFD. $C'$ represents the indicators of *Slope* and *Distance from streams* while $C''$ represents the rest factors. $\omega'_{pj}$ and the $\omega''_{pi}$ are the responding weight elements of $C'$ and $C''$, respectively.

The flood risk maps were produced via the flood risk indexes using the Natural Break method. In this study, we derived the flood maps by slicing the flood risk index into five classes, respectively labeled as "*Very Low*" (class 1), "*Low*" (class 2), "*Normal*" (class 3), "*High*" (class 4), and "*Very High*" (class 5).

### *3.6. Validating flood risk estimation results*

In this study, we adopted the correct ratio and fit ratio to assess flood estimation accuracy, following (Zhang et al., 2021), (Alfieri et al., 2014; Bates and De Roo, 2000):

$$Correct(\%) = \frac{FA_{FRI} \cap FA_{Water}}{FA_{Water}} \times 100 \qquad (11)$$

$$Fit(\%) = \frac{FA_{FRI} \cap FA_{Water}}{FA_{FRI} \cup FA_{Water}} \times 100 \qquad (12)$$

where $Correct(\%)$ is the correct ratio, and $Fit(\%)$ is the fit ratio. $FA_{FRI}$ suggests areas with a high likelihood of being flooded. In this study, we construct two validation sets as: {"*High*", "*Very high*"}. $FA_{Water}$ represents the flood cells extracted from the GF-3 and Landsat 8 OLI images.

## 4. Results

In this session, we described the experimental results and discussed our findings. We analyzed the flood risk sensitivity among models and among judgment matrixes definitions. To explore the different sensitivity from models, we calculated the correct ratio and fit ratio of models and analyzed the statistical characteristics of the flood risk levels distribution via judgment matrixes definitions. To distinguish the inner differences within models, we discussed the flood risk estimation by a certain cluster of judgment matrixes definition. We further discussed the judgement matrixes influence in different type of basins. This section is organized as follows: Subsection 4.1 compared the model performance via custom judgment matrixes from 48 projects. Subsection 4.2 compared the differences in using the 48 judgment matrix definitions derived flood risk level distribution in typical basins. Subsection 4.3 discussed the influence of judgment matrixes in flood risk level distribution.

### 4.1. The performance differences among models

To compare the influence of different criteria weight in five used AHP related models (i.e., pixel-AHP, WZSAHP of MFD-RC, WZSAHP of MFD-All, WZSAHP of D8-RC, WZSAHP of D8-All), we calculated their correct ratio and fit ratio by setting estimated flood risk levels "*High*" and "*Very High*" pixels as *True* and "*Very Low*", "*Low*", and "*Normal*" as False, and compared them with the ground-truthing flooded areas in the validation dataset, as shown in Table.4. Note that permanent water areas were excluded in the validation dataset following China's impervious surface grid product (2m) (Shao et al., 2019). The remote sensing images were classified as *water* and *none water* cells. And the ground-truthing *water* cells were labeled as *Ture*, while the ground-truthing *none-water* cells were labeled as *False*.

Table 7. The correct ratio and the fit ratio of five investigated models in 48 projects.

| Prj. No. | S/E | S/R | E/R | AHP(%) | | MFD-RC(%) | | MFD-All(%) | | D8-RC(%) | | D8-All(%) | |
|---|---|---|---|---|---|---|---|---|---|---|---|---|---|
| | | | | Cor. | Fit | Cor. | Fit | Cor. | Fit | Cor. | Fit | Cor. | Fit |
| 1 | 4 | 1/2 | 1/3 | 23 | 12 | 59 | 18 | 23 | 12 | 44 | 13 | 85 | 12 |
| 2 | 4 | 1/2 | 1/4 | 19 | 12 | 59 | 18 | 19 | 12 | 51 | 13 | 85 | 13 |
| 3 | 4 | 1/2 | 1/5 | 22 | 12 | 59 | 18 | 22 | 12 | 46 | 13 | 85 | 12 |
| 4 | 4 | 1/2 | 1/6 | 22 | 13 | 59 | 18 | 22 | 13 | 46 | 13 | 85 | 12 |
| 5 | 4 | 1/3 | 1/3 | 16 | 11 | 59 | 18 | 16 | 11 | 47 | 13 | 85 | 12 |
| 6 | 4 | 1/3 | 1/4 | 18 | 12 | 59 | 18 | 18 | 12 | 49 | 13 | 86 | 11 |
| 7 | 4 | 1/3 | 1/5 | 18 | 11 | 59 | 18 | 18 | 11 | 49 | 13 | 86 | 11 |
| 8 | 4 | 1/3 | 1/6 | 16 | 11 | 59 | 18 | 16 | 11 | 46 | 13 | 86 | 12 |
| 9 | 5 | 1/2 | 1/3 | 23 | 12 | 59 | 18 | 23 | 12 | 44 | 13 | 85 | 12 |
| 10 | 5 | 1/2 | 1/4 | 22 | 12 | 59 | 18 | 22 | 12 | 47 | 13 | 85 | 12 |
| 11 | 5 | 1/2 | 1/5 | 21 | 12 | 59 | 18 | 21 | 12 | 46 | 13 | 86 | 11 |
| 12 | 5 | 1/2 | 1/6 | 21 | 12 | 59 | 18 | 21 | 12 | 46 | 13 | 85 | 12 |
| 13 | 5 | 1/3 | 1/3 | 22 | 13 | 59 | 18 | 22 | 13 | 49 | 13 | 85 | 12 |
| 14 | 5 | 1/3 | 1/4 | 21 | 13 | 59 | 18 | 21 | 13 | 49 | 13 | 85 | 12 |
| 15 | 5 | 1/3 | 1/5 | 18 | 11 | 59 | 18 | 18 | 11 | 46 | 13 | 85 | 12 |
| 16 | 5 | 1/3 | 1/6 | 18 | 12 | 59 | 18 | 18 | 12 | 46 | 13 | 86 | 12 |
| 17 | 6 | 1/2 | 1/3 | 22 | 12 | 60 | 17 | 22 | 12 | 49 | 13 | 85 | 12 |
| 18 | 6 | 1/2 | 1/4 | 21 | 12 | 59 | 18 | 21 | 12 | 47 | 13 | 85 | 13 |
| 19 | 6 | 1/2 | 1/5 | 16 | 10 | 60 | 18 | 16 | 10 | 46 | 13 | 85 | 13 |
| 20 | 6 | 1/2 | 1/6 | 22 | 13 | 59 | 18 | 22 | 13 | 46 | 12 | 85 | 12 |
| 21 | 6 | 1/3 | 1/3 | 23 | 13 | 60 | 18 | 23 | 13 | 49 | 13 | 85 | 12 |
| 22 | 6 | 1/3 | 1/4 | 21 | 13 | 59 | 18 | 21 | 13 | 49 | 13 | 85 | 12 |
| 23 | 6 | 1/3 | 1/5 | 17 | 11 | 59 | 18 | 17 | 11 | 49 | 13 | 86 | 11 |
| 24 | 6 | 1/3 | 1/6 | 18 | 11 | 59 | 18 | 18 | 11 | 49 | 13 | 86 | 12 |
| 25 | 7 | 1/2 | 1/3 | 22 | 12 | 60 | 17 | 22 | 12 | 47 | 13 | 85 | 12 |
| 26 | 7 | 1/2 | 1/4 | 22 | 12 | 60 | 18 | 22 | 12 | 47 | 13 | 85 | 12 |
| 27 | 7 | 1/2 | 1/5 | 22 | 12 | 60 | 17 | 22 | 12 | 46 | 12 | 86 | 11 |
| 28 | 7 | 1/2 | 1/6 | 21 | 12 | 60 | 18 | 21 | 12 | 46 | 12 | 85 | 12 |
| 29 | 7 | 1/3 | 1/3 | 23 | 13 | 60 | 18 | 23 | 13 | 49 | 13 | 85 | 12 |
| 30 | 7 | 1/3 | 1/4 | 21 | 12 | 59 | 18 | 21 | 12 | 49 | 13 | 85 | 12 |
| 31 | 7 | 1/3 | 1/5 | 21 | 12 | 59 | 18 | 21 | 12 | 46 | 13 | 85 | 12 |
| 32 | 7 | 1/3 | 1/6 | 18 | 11 | 59 | 18 | 18 | 11 | 49 | 13 | 85 | 13 |
| 33 | 8 | 1/2 | 1/3 | 22 | 12 | 60 | 17 | 22 | 12 | 44 | 13 | 85 | 12 |
| 34 | 8 | 1/2 | 1/4 | 23 | 12 | 60 | 18 | 23 | 12 | 51 | 13 | 85 | 13 |
| 35 | 8 | 1/2 | 1/5 | 22 | 12 | 60 | 17 | 22 | 12 | 46 | 12 | 85 | 12 |
| 36 | 8 | 1/2 | 1/6 | 19 | 12 | 60 | 18 | 19 | 12 | 46 | 12 | 85 | 12 |
| 37 | 8 | 1/3 | 1/3 | 22 | 13 | 60 | 18 | 22 | 13 | 49 | 13 | 85 | 12 |
| 38 | 8 | 1/3 | 1/4 | 20 | 12 | 60 | 17 | 20 | 12 | 49 | 13 | 85 | 12 |
| 39 | 8 | 1/3 | 1/5 | 20 | 12 | 60 | 18 | 20 | 12 | 46 | 13 | 85 | 12 |
| 40 | 8 | 1/3 | 1/6 | 20 | 12 | 60 | 18 | 20 | 12 | 49 | 13 | 86 | 12 |
| 41 | 9 | 1/2 | 1/3 | 22 | 12 | 60 | 17 | 22 | 12 | 47 | 13 | 85 | 13 |
| 42 | 9 | 1/2 | 1/4 | 22 | 12 | 60 | 18 | 22 | 12 | 47 | 13 | 85 | 13 |
| 43 | 9 | 1/2 | 1/5 | 22 | 12 | 60 | 17 | 22 | 12 | 46 | 12 | 85 | 12 |
| 44 | 9 | 1/2 | 1/6 | 22 | 12 | 60 | 17 | 22 | 12 | 46 | 12 | 85 | 12 |
| 45 | 9 | 1/3 | 1/3 | 22 | 13 | 60 | 18 | 22 | 13 | 49 | 13 | 85 | 12 |
| 46 | 9 | 1/3 | 1/4 | 22 | 13 | 59 | 18 | 22 | 13 | 49 | 13 | 85 | 12 |
| 47 | 9 | 1/3 | 1/5 | 21 | 12 | 60 | 18 | 21 | 12 | 49 | 13 | 85 | 13 |
| 48 | 9 | 1/3 | 1/6 | 20 | 12 | 60 | 18 | 20 | 12 | 49 | 13 | 86 | 12 |

In Table 4, the column name "*Prj. No.*" represents the project number. The column name "*S/E*", "*S/R*", and "*E/R*" suggest the pairwise element name of the judgment matrix, representing (*Slope*, *Elevation*), (*Slope*, *Distance from streams*), and (*Elevation*, *Distance from streams*). The column names of "*cor.*" represents the correct ratio. From the value distribution of the correct ratio and fit ratio in the 48 projects, we notice that the correct ratio and fit ratio of WZSAHP MFD-RC are mostly {59, 60} and {17, 18}. The correct ratio and fit ratio of D8-All are mostly {85, 86} and {11, 12, 13}. For pixel-AHP, WZSAHP MFD-All, WZSAHP MFD-RC, WZSAHP D8-All, and WZSAHP D8-RC models, their correct ratio and fit ratio fluctuate irregularly with changing judgment matrix, as shown in Fig.6.

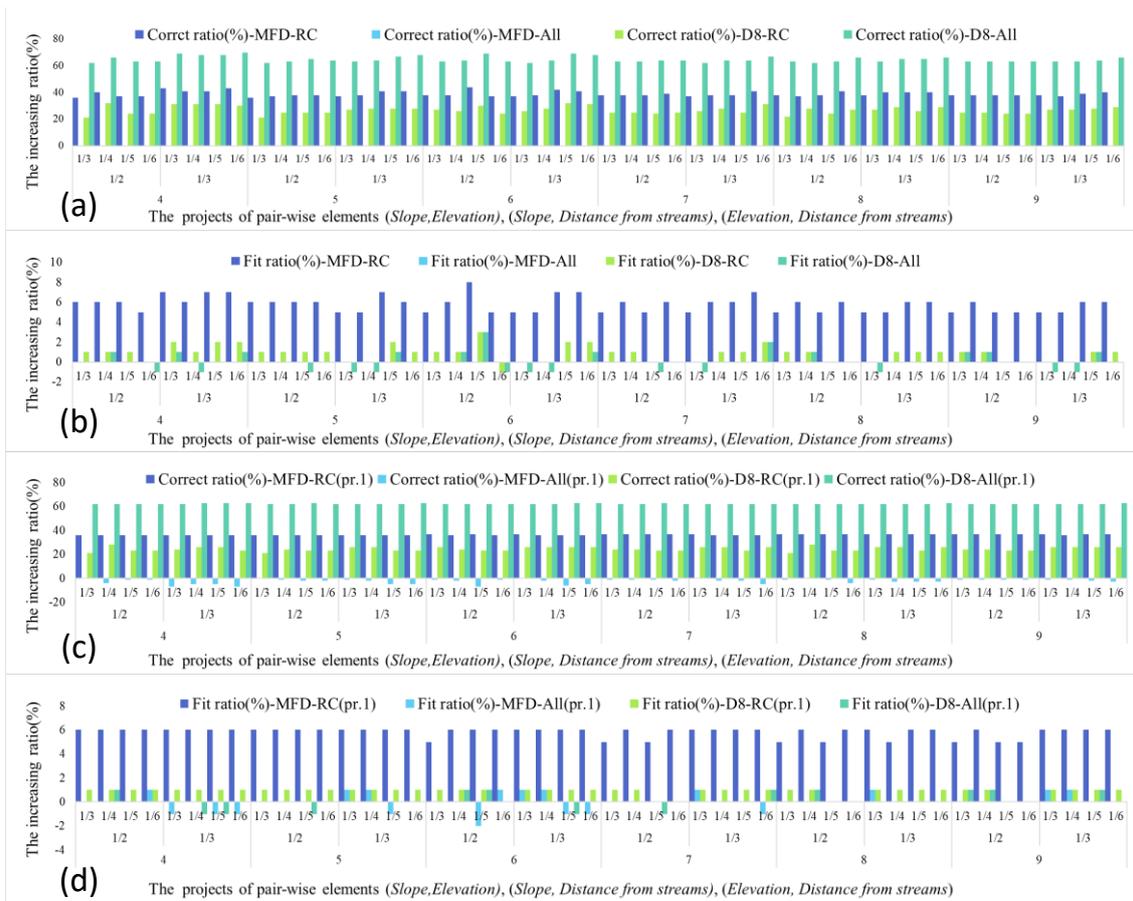

Figure 6 Performance comparison of changing criteria definitions between the pixel-AHP with the four WZSAHP related models. Fig.(a)-(d) show the performance differences using the WZSAHP models and pixel-AHP. The criteria definitions of the comparing WZSAHP and pixel-AHP models are the same as in Fig.(a) and Fig.(b), while the definitions of Fig.(c) and (d) are set following project1.

Comparing with the pixel-AHP, we notice increased correct ratios and fit ratios for WZSAHP models in Fig.6. As shown in Fig.6(a) and Fig.6(b), all the correct ratios of the MFD-RC, the MFD-All, the D8-RC, and the D8-All are higher than the pixel-AHP model. We also notice that the fit ratio of MFD-RC is always higher than pixel-AHP. But in some criteria definitions, the fit ratio of MFD-All, D8-RC, and D8-All can be lower than the responding pixel-AHP. As shown in Fig.6(c) and Fig.6(d), by setting pixel-AHP criteria definition as project1, the MFD-RC, the D8-RC and the D8-All achieve increased correct ratios and fit ratios, while for the MFD-All model, correct ratios and fit ratios of certain definitions are lower comparing to AHP models. Therefore, it can be concluded that the MFD-RC model is able to ensure improvements of correct ratio and fit ratio using all the criteria definitions in this study, while we do not see a steady increase in the correct ratio and fit ratio in other models.

### 4.2. The influence of pairwise definitions in WZSAHP-RC(MFD)

In this subsection, we further discuss differences in flood risk levels caused by the definition of the pairwise element in the judgment matrixes among different types of basins. The typical basins were chosen as two groups, contain outlets and not contain outlest. As the main outlets of the study area are Chao Lake and the Yangtze River, as shown in Fig.7. Therefore, three basins were chosen for analysis, respectively denoted as *Basin1*, *Basin2*, and *Basin3*.

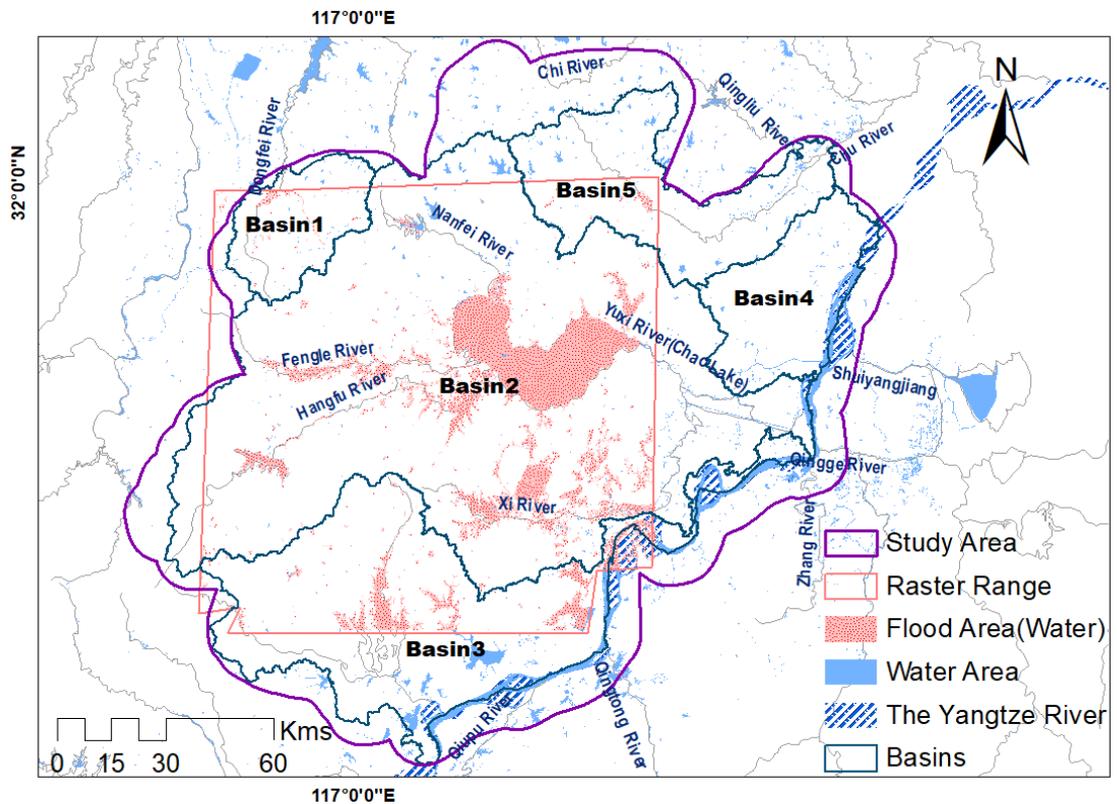

Figure 7. The distribution of five basins in the study area.

As shown in Fig.7, *Basin1* contains no outlet, meaning that the collected rainwater flows away from the study area. *Basin2* contains Chao Lake and the Yangtze River, it has a large area of permanent water, and its flood risk is mainly brought by the Yangtze River and the Chao Lake. *Basin3* is located on the north side of the Yangtze River.

Taking *Basin1*, *Basin2*, and *Bains3* as mask layers, we extracted the distribution of flood risk levels from the WZSAHP MFD-RC model using the 48 defined criteria matrixes, as shown in Fig.8, Fig.9, and Fig.10. Sub-figures were labeled related to the project number shown in Table.4. The flood risk maps are designed following a five-color schema, i.e., {*Green*, *Light Green*, *Yellow*, *Orange*, *Red*} to map flood risk levels of {*Very Low*, *Low*, *Normal*, *High*, *Very High*}, respectively.

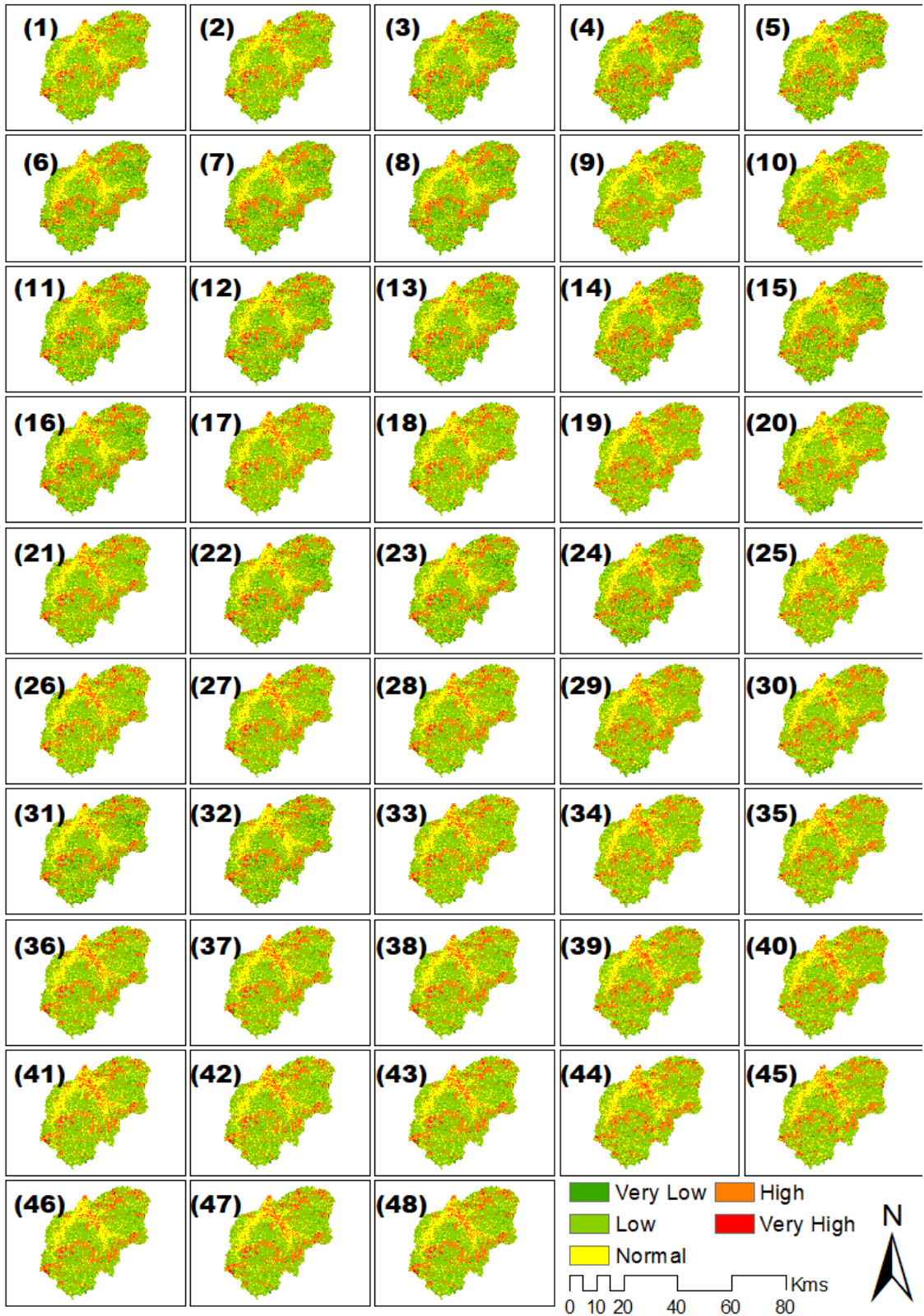

Figure 8. The distribution of flood risk levels of *Basin1* derived by the MFD-RC via WZSAHP.

As shown in Fig.8, flood risk levels of *"Very Low"*, *"Low"* and *"Normal"* are dominant in most of the areas, and the flood risk levels from all 48 projects are rather similar. All the figures suggest that areas with risk levels of *"High"* and *"Very high"* are distributed in the low-lying areas in *Basin1*, while most of the areas in and nearby the Dongfei River are in the risk level of *"Normal"*. The flood risk levels of *"High"* and *"Very High"* are distributed in the north, in the middle of the north to south, and at the mid-bottom of the basin.

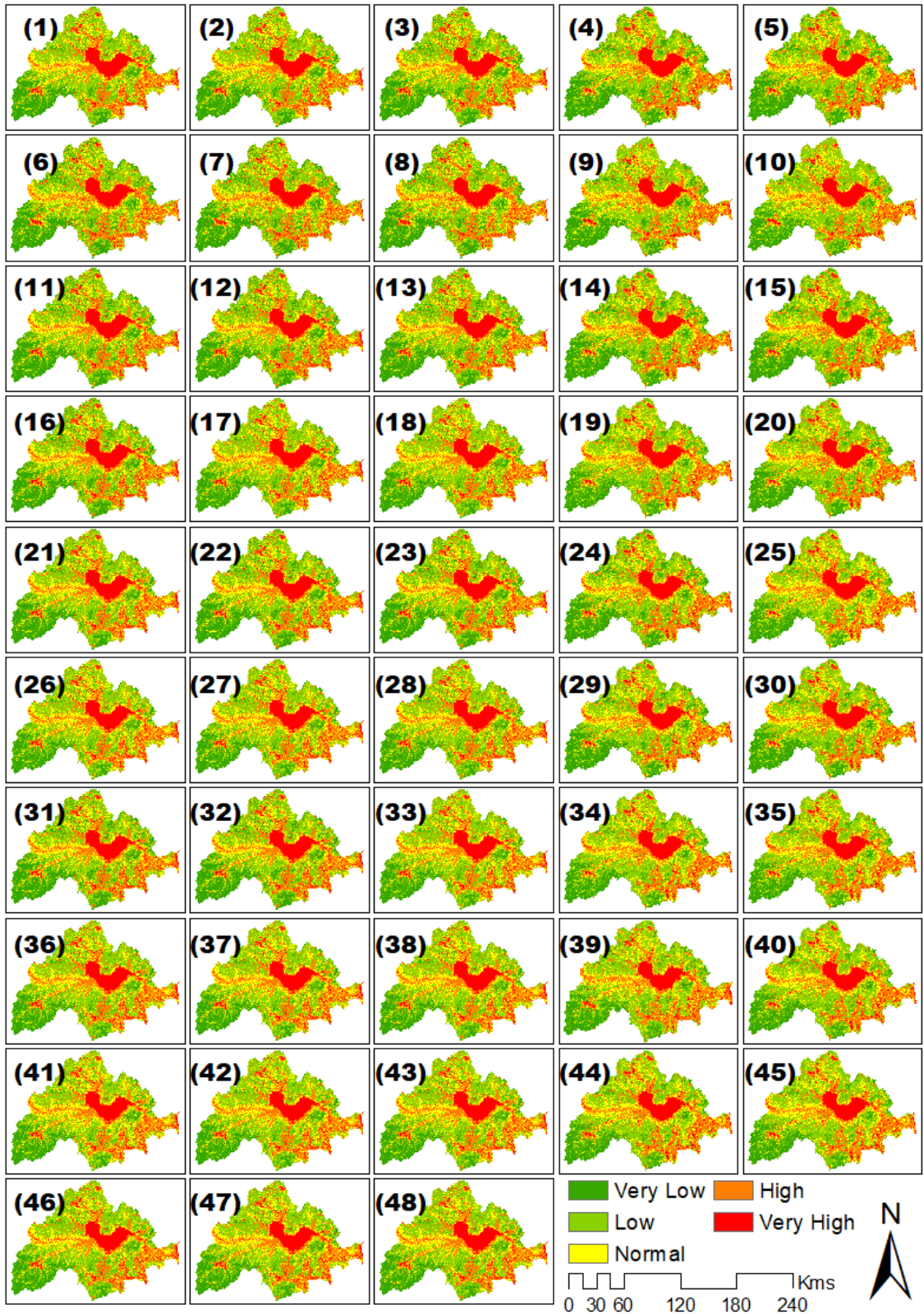

Figure 9. The distribution of flood risk levels of *Basin2* derived by the MFD-RC via WZSAHP.

As shown in Fig.9, areas with flood risk levels of *"High"* and *"Very high"* in *Basin2* are considerably more extensive compared to *Basin1*. The distribution of flood risk levels of the 48 projects is similar. Areas with flood risk levels of *"High"* and *"Very High"* are distributed in the range of water systems around Chao Lake, including the areas nearby the Nanfei River, the Fengle River, the Hangfu River, the Zhao River, the Xi River, and the Yuxi River.

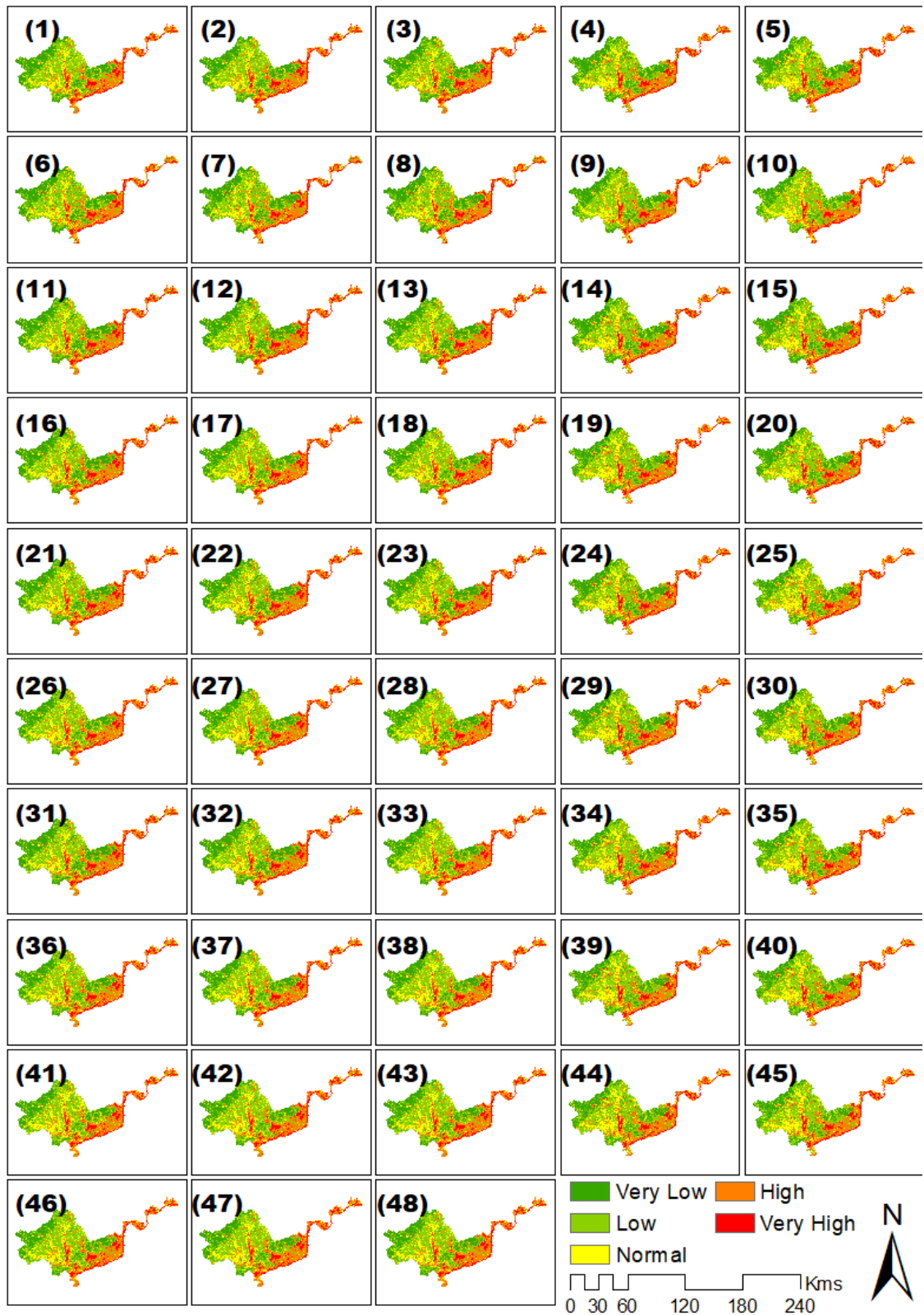

Figure 10. The distribution of flood risk levels in *Basin3*, derived by the MFD-RC via WZSAHP.

As shown in Fig.10, the distribution of flood risk levels of the 48 projects in *Basin3* also presents similar patterns. Areas with flood risk levels of "*High*" and "*Very High*" are distributed nearby the Yangtze River. The risk level of the reservoirs and small lakes areas generally is "*Normal*". According to the average distribution of flood risk levels, we further cumulated the quality of pixels grouped by flood risk levels, as shown in Fig.11.

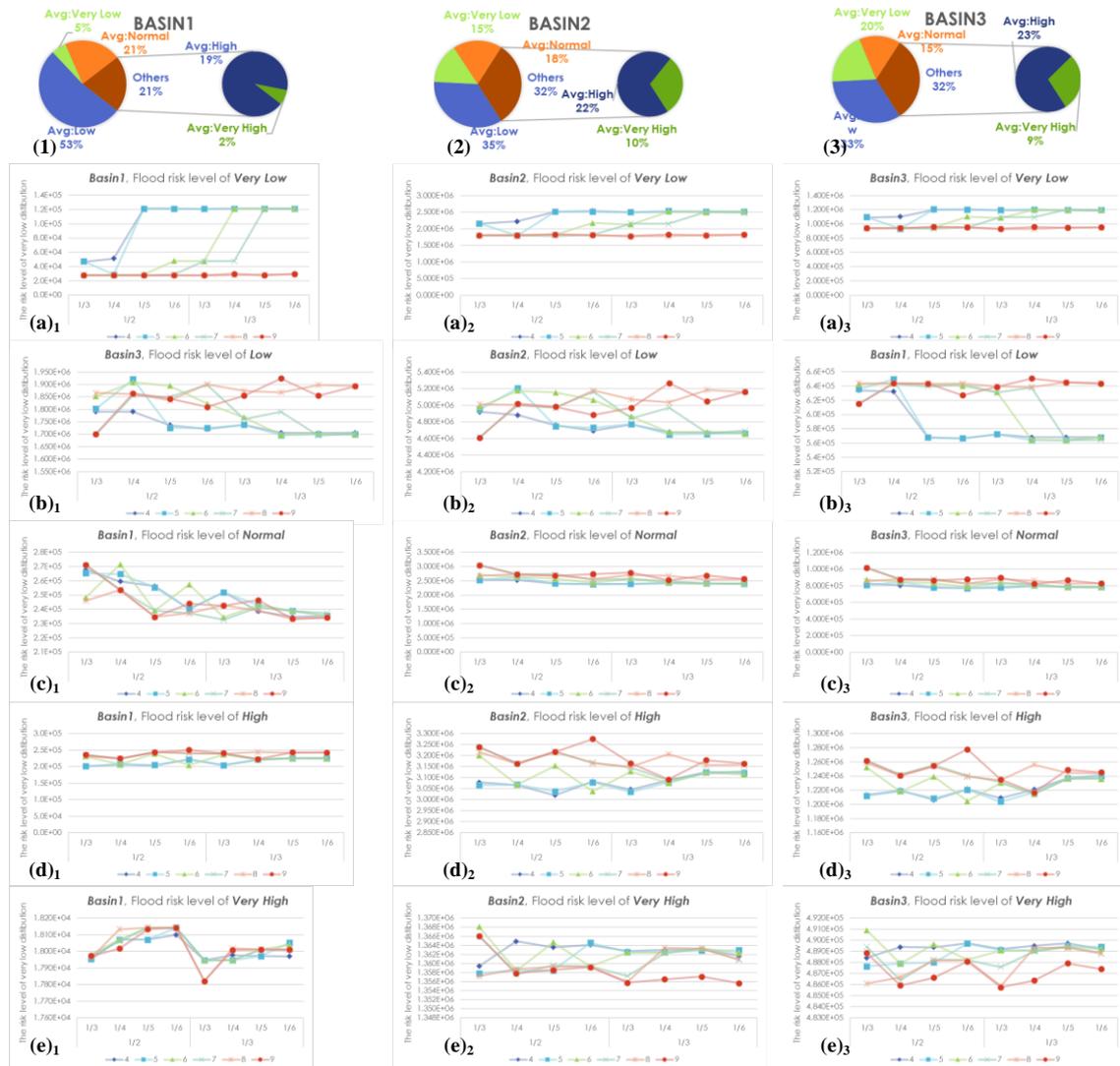

Figure 11. The quality of pixels grouped by flood risk levels in *Basin1*.

Fig.11(1)-(3) show the statistical pie chart of the average areas grouped by flood risk levels from the 48 projects. The area ratio order of the "*High*" and "*Very High*" follows: *Basin2(32%)* =

*Basin3(32%) > Basin1(21%)*, the order of *Normal* follows: *Basin1(21%) > Basin2(18%) > Basin3(15%)*, and the order of "*Low*" and "*Very Low*" follows: *Basin1(58%) > Basin3(53%) > Basin2(50%)*.

Fig.11 (a)-(e) present the distribution of flood risk levels from all 48 projects grouped by the pairwise element definition of (*Slope*, *Elevation*). Fig.11 (a)$_1$-(e)$_1$, Fig.13 (a)$_2$-(e)$_2$, and Fig.11 (a)$_3$-(e)$_3$ correspond to *Basin1*, *Basin2*, and *Basin3*, respectively. We notice that *Basin2* and *Basin3* have similar distribution patterns of flood risk levels influenced by the definition of pairwise elements in judgment matrixes. For *Basin2* and *Basin3*, under the condition of (*Slope*, *Elevation*) = {8, 9}, areas (the (*Slope*, *Elevation*) = {8, 9}) with flood risks of "*Very Low*" and "*Very High*" are lower while areas with flood risk of Normal are considerably higher than the condition of (*Slope*, *Elevation*) = {5, 6, 7}. However, this regular pattern is not presented in *Basin1*.

### *4.3. The estimated flood risk levels among investigated methods*

In this study, we selected a total of 48 definitions of judgment matrix (i.e., projects) as comparing criteria. Five models were investigated in this study, including the pixel-AHP, the WZSAHP models using sub-watersheds delimitated by MFD to constraint related indicators (MFD-RC) and the whole indicators (MFD-All), and the WZSAHP-RC models using sub-watersheds delimitated by D8 to constraint related indicators (D8-RC) and the whole indicators (D8-All). Fig. 12 presents the derived flood risk levels, i.e., "*Very Low*", "*Low*", "*Normal*", "*High*", "*Very High*", from these models.

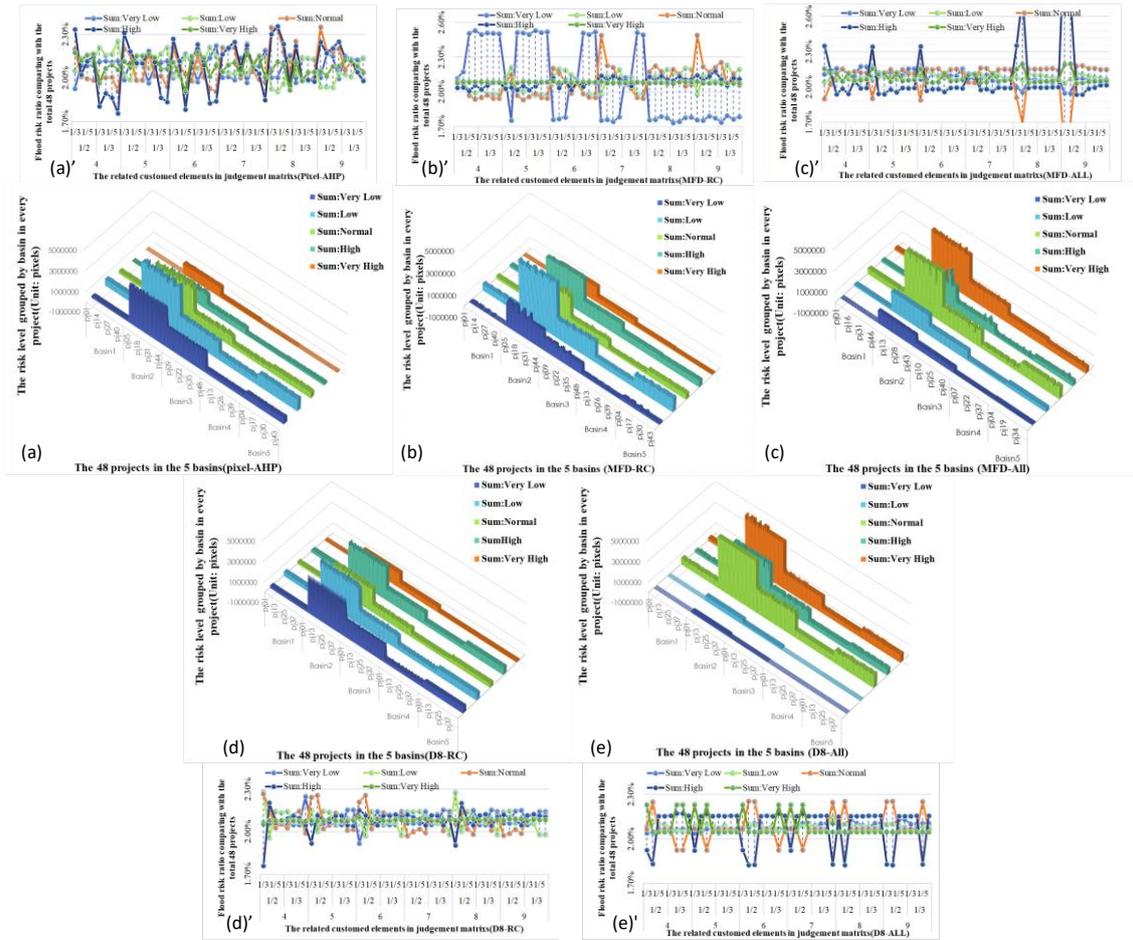

Figure 12. The distribution of flood risk levels from the 48 projects. Figures (a)-(e) present the histograms of the five flood risk levels, while Figures (a)'-(e)' present the ratios of flood risk levels in the five basins of each project.

The five flood risk levels calculated at the basin level in each project are drawn as a perspective view of 3D histograms, shown in Fig.12(a)-(e). The X-axis denotes project numbers in each basin. We notice that the distribution of flood risk levels from the pixel-AHP model change considerably among projects (Fig.12(a)). For the MFD-RC model (Fig.12(b)) and the D8-RC model (Fig.12(d)), "*Low*", "*Very Low*", and "*Normal*" pixels are distributed in a different manner among projects, while "*High*" and "*Very High*" pixels are with a rather similar distribution pattern. Comparing the MFD-All model (Fig.12(c)) with the D8-All model (Fig.12(e)), for *Basin1*, *Basin3*, *Basin4*, and *Baisn5*, distribution patterns of "*High*" and "*Very High*" greatly differ among projects. In comparison, for *Basin2*, risk levels of "*Low*", "*Very Low*", and "*Normal*" remain stable.

33
Submmited to pre-review

Fig.12(a)′-(e)′ show the relative changes among the investigated projects in five selected models. The X-axis denotes the project number, while the Y-axis denotes the flood risk level ratio. For pixel-AHP (Fig.12(a)′), the derived flood risk level ratios change in an irregular pattern. In comparison, for other four models, i.e., MFD-RC(Fig.12(b)′), MFD-All Fig.12(c)′, D8-RC (Fig.12(d)′), D8-All (Fig.12 (e)′), the variation of flood risk level ratios fluctuate less than than the pixel-AHP model. Especially for the MFD-RC model (Fig.12(b)′), despite that flood risk ratios of "*Very Low*", "*Low*", and "*Normal*" present changing patterns, ratios of "*High*" and "*Very High*" remain steady. For the MFD-All model (Fig.12(c)′), the "*High*" and "*Normal*" fluctuate around the expected mean value. When (*Slope*, *Elevation*) = {4, 5, 6, 8, 9}, (*Slope*, *Distance from streams*) = 1/2, and (*Slope*, *Distance from streams*) = 1/3, the areas with the flood risk of "*Very High*" increases sharply. For the D8-RC model (Fig.12(d)′), considerable fluctuations can be observed for more than 11 projects. In comparison, for the D8-All model (Fig.12(e)′), irregular fluctuations can be observed for more than 13 projects.

The above results indicate that the pixel-AHP is sensitive to weight definitions, and the flood risks of "*High*" and "*Very High*" are considerably steady in the MFD-RC model.

## *5. Discussion*

AHP models are widely used in flood risk estimation. Numerous studies have shown that the definition of weight vectors calculated by the judgment matrix largely determines the derived flood risk levels (Bathrellos et al., 2017; Huu et al., 2020; Kanani-Sadat et al., 2019; Skilodimou et al., 2019). The definition of pairwise elements of the judgment matrix is the essential step in flood risk estimation. For pixel-AHP models that use pixels as the basic unit, even the changes in criteria are tiny, the flood risk indexes tend to vary, resulting in great spatial heterogeneity. This explains why the flood risk estimation via pixel-AHP shows great sensitivity to the changes of the judgment matrix. Thus, other options beyond pixel-wise analysis need to be investigated.

According to our former work, considering sub-watershed as a unit to constrain all indicators via the maximum statistical value, i.e., WZSAHP, and to constrain only converging related indicators. i.e., MFD-RC, both demonstrate better performances compared to pixel-AHP (Zhang et al., 2021). Thus, we focus on investigating the sensitivity of judgment matrix definition on flood risk estimation in this study. Following the importance order of *Distance from streams > slope > Elevation*, we constructed a total of 48 judgment matrixes and compared the performance of the WZSAHP model that includes MFD-RC, MFD-All, D8-RC, D8-All, with a pixel-AHP model. We found that comparing with the pixel-AHP model, all WZSAHP models fluctuated less given different definitions of the judgment matrix. Specifically, the MFD-RC model achieved stable increases in the correct ratio and the fit ratio compared to MFD-All, D8-RC, and D8-All models. Since the MFD algorithm identifies sub-watersheds using pixels that flow in the same sink, while the D8 algorithm identifies sub-watersheds using pixels that flow out from the same outlet, constraint using the MFD algorithm is a more suitable approach that approximates the natural flooding process than the D8 algorithm. This may be the reason why the models using MFD delimitated sub-watersheds outperform models using D8 delimitated sub-watersheds.

*6. Conclusions*

In this study, we evaluated and discussed the sensitivity of flood risk estimation brought by judgment matrix definition by investigating the performance of pixel-based and sub-watershed-based AHP models. While the pixel-AHP model estimates flood risk using pixel-wise indicators, sub-watershed-based AHP models adopt sub-watershed as the basic unit to constrain indicators via the maximum statistical value (WZSAHP) (Zhang et al., 2021). In this study, we adopted a total of 48 different judgment matrixes and further estimated flood risk levels via the pixel-AHP and the WZSAHP model that includes MFD-RC, MFD-All, D8-RC, and D8-All. Validating against the flooded areas extracted from remote sensing images, we notice that the MFD-RC model achieved steady increases in both the correct ratio and the fit ratio compared with pixel-AHP, MFD-All, D8-RC, and D8-All models. The results also suggested that the performance of the pixel-AHP model fluctuates intensively given different definitions of judgment matrixes, while WZSAHP models fluctuate considerably less than the pixel-AHP model. We believe there is still room for further investigation and discussion on the internal mechanism.

*Acknowledgments*

This work was supported in part by the National Key R&D Program of China [grant 2018YFB2100501], the Fundamental Research Funds for the Central Universities [grant number 2042021kf0007], and the National Natural Science Foundation of China [grant numbers 42090012, 41771452, 41771454 and 41901340]. The authors would like to thank the anonymous reviewers and editors for their comments, which helped us improve this paper significantly.

*Appendix A. Supplementary data*

Supplementary data associated with this work can be found in the online version. These data include the rating dataset of involved indicators for flood risk estimation, the watersheds delimited by D8 and MFD algorithms (with an area threshold of 66.7 ha), and the boundaries of water bodies.

Table A.1 The details of supplementary data.

| Data name | Data type | Detail | Description |
|---|---|---|---|
| materials.gdb | ARCGIS GDB | R_DEM_5_r_n | The ranked elevation indicator. |
| | | R_slope_5_r_n | The ranked slope indicator. |
| | | R_IMP | The ranked Hydro-lithological formations indicator. |
| | | R_LANDUSE | The ranked Land use type indicator. |
| | | R_ED_Stram_water | The ranked Distance from streams indicator. |
| | | Watershed_D866.7 | The watershed derived by the D8 algorithm, with area thresholds of 66.7 ha. |
| | | Watershed_MFD | The watershed derived by the MFD algorithm. |

| | | Flood_water_body | The water body extracted during the flooding period. |
| --- | --- | --- | --- |
| | | Mask_water_body | Normal water body extracted in non-flooding days. |
| submit-detailCellStatistics.xlsx | EXCEL | | The flood risk accuracy calculation results and the detail flood risk level distribution corresponding to flood or non-flood cells. |

*Appendix B. The indicator pairwise definition in the judgment matrix*

According to the pairwise elements of (Slope, Elevation), (Slope, Distance from Stream), (Elevation, Distance from stream) defined in judgment matrixes, and the other related custom elements in the judgment matrixes, the final weight vectors in the 48 projects can be calculated. Their corresponding rose figures were presented in Fig.B.1.

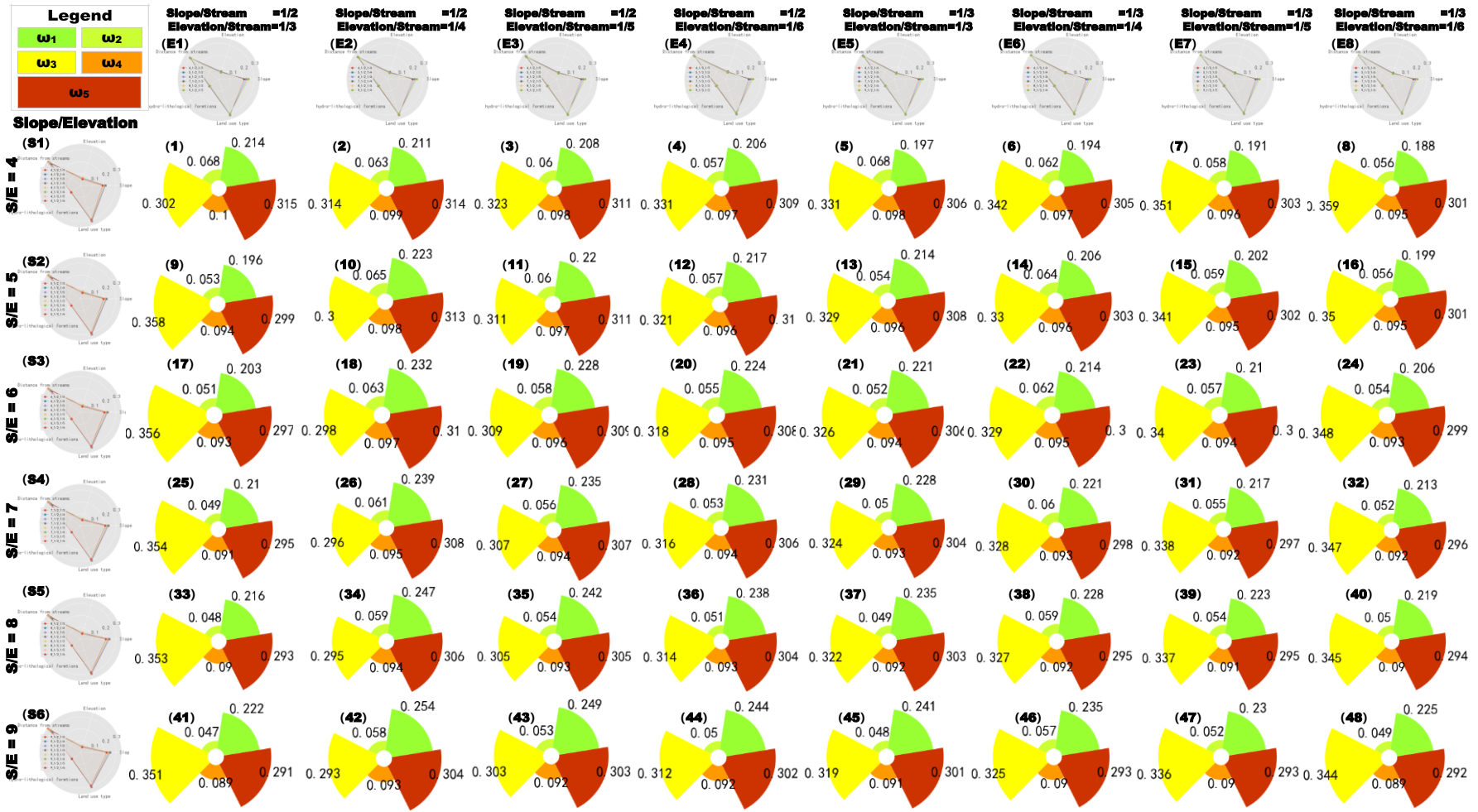

Figure B.1. The rose plots of weight vectors for all 48 projects. Fig. (1)-(48) present rose plots that correspond to projects (1)-(48). Fig. (S1)-(S6) are the radar plots grouped by judgment elements of (*Slope*, *Elevation*) in a set of {4, 5, 6, 7, 8, 9}. Fig. (E1)-E(8) are the radar plots grouped by judgment elements of (*Slope*, *Distance from streams*) in a set of {1/2, 1/3} coupled with (*Elevation*, *Distance from streams)* in a set of {1/3, 1/4, 1/5, 1/6}.